\documentstyle[epsf,12pt]{article}
\begin{document}

\begin{titlepage}
\begin{flushright}
IFUP--TH/2008-37\\
\end{flushright}
~

\vskip .8truecm
\begin{center}
\Large\bf
Consistent gravitational anomalies for chiral bosons
\footnote{
Work supported in part by M.U.I.R.}
\end{center}

\vskip 1.6truecm
\begin{center}
{Stefano Giaccari and Pietro Menotti} \\
\vskip .8truecm
{\small\it Dipartimento di Fisica, Universit{\`a} di Pisa and\\
INFN, Sezione di Pisa, Largo B. Pontecorvo 3, I-56127}\\
\end{center}
                                                                               
\vskip 1.2truecm
               
\vskip 1.2truecm
                                                              
\begin{abstract}

Exact consistent gravitational anomalies for chiral bosons in two dimensions
are treated both with the Schwinger-DeWitt regularization and independently
through a cohomological procedure. The diffeomorphism transformations are
described by a single ghost which allows to climb the cohomological chain in a
unique way.

\end{abstract}

\vskip 1truecm

\end{titlepage}

\section{Introduction}

Anomalies play a very important role in quantum field theory both at the
phenomenological and at the fundamental level \cite{bertlmann}. 
Gravitational
anomalies were discovered somewhat later \cite{AGW} 
probably due to the fact
that pure gravitational anomalies do not exist in four dimensional space but
only in dimension $2+4n$. The simplest instance of pure gravitational anomaly
is the one due to the presence of a chiral fermion in two dimensions
\cite{AGW,leutwyler,BK}.
There are also gravitational anomalies produced by boson fields i.e. by the
self-dual and anti self-dual fields which are realized in the simplest instance
by the chiral bosons in dimension $2$. The coupling of (anti) self-dual 
tensors to gravity was given by Henneaux and Teitelboim 
\cite{HT,HTcile}. 
The formulation is not explicitely covariant even if it satisfies all the
requirements under diffeomorphism transformations. In the case of the chiral
boson in two dimensions the Henneaux-Teitelboim action extends the
Floreanini-Jackiw \cite{FJ} action on flat space to the presence of an
external  gravitational field. Perturbative anomalies for (anti)
self-dual 
tensor fields were computed in \cite{AGW,BvNpert,BvNbook} and
exact covariant gravitational anomalies were computed in \cite{BvNbook}.

Powerful cohomological methods were developed in the literature
\cite{zumino}-\cite{BTvP2} 
which in the most developed approach in the gravitational case
\cite{BDK3} exploit the properties of Lorentz transformations embodied by the
presence of the vierbeins. The general case of scalar matter conformally
coupled with gravity was treated in \cite{BTvP2}.

In the present paper we give a simple self contained derivation of the
consistent gravitational anomaly for chiral bosons in two dimensions
starting from the Henneaux-Teitelboim action. Only one 
function $K$ of the metric enters the action. Becchi \cite{becchi} was the
first to point out that in each chiral sector of conformally invariant two
dimensional field theory the metric enters through a single function and the
diffeomorphisms are described by a single ghost.  
Even if the mentioned function $K$ which summarizes
the gravitational field can be written as the
ratio of two zweibein components, such a ratio is exactly invariant under
local Lorentz transformations and thus the local Lorentz group leaves no trace
in the action.
The problem will be solved
in two independent ways, the first by using conventional field theory
techniques and the other exploiting cohomological methods. The first approach
is near in spirit, although technically different,  to the exact treatment of 
the gravitational anomaly given by Leutwyler \cite{leutwyler} and Leutwyler 
and Mallik
\cite{LM} for the chiral fermion case
in two dimensions as both use the Schwinger-DeWitt \cite{schwinger,dewitt}
regularization of the
functional integral. The cohomological method which follows in Section
\ref{cohosection} is
completely independent of the previous one and owes its simplicity to the 
fact that
in the case of the chiral bosons in two dimensions only one function $K$ of
the metric enters 
and that the diffeomorphisms are represented just by one ghost \cite{becchi}. 
It is very
simple to write down the descending cohomological chain and prove from the
last term, that the found anomaly is non trivial. It is of interest that
the final term, of ghost number $3$, is unique up to trivial
additions and that the cohomological chain can be climbed up in a unique way,
finding the result already obtained with the functional integral. For doing
that one has to consider three sequences; of these two are exact while the
other is exact but for a junction, where the anomaly arises.
 
The paper is structured as follows. In Section 2 we lay down the formalism and
discuss some properties of the Henneaux-Teitelboim action.  In Section 3 we
show how the anomaly at the perturbative level i.e. for weak gravitational
field, can be easily found by exploiting the standard Pauli-Villars (PV)
treatment of the non chiral bosons and we compare the result with the one
obtained in the light-cone treatment \cite{BvNbook}. The PV method can be
applied also in dimensions greater than $2$.
In Section 4 we give the non perturbative calculation of the anomaly by using
the Schwinger-DeWitt method.
In Section 5 we prove that the found non perturbative anomaly is consistent
i.e. satisfies the Wess-Zumino relation, and we give an algebraically proof
that the found anomaly is non trivial.
In Section 6 we give the cohomological treatment of the anomaly problem
deriving the results obtained in the previous sections in a purely algebraic
way. First one shows the uniqueness of the last term in the cohomological
chain and then one shows that three cohomological sequences are exact but for
one junction in a sequence which gives rise to the anomaly. Such a treatment
is self contained and 
completely independent from the ones given in Section 3 and Section 4.
In Section 7 we give some concluding remarks. In Appendix A and B we elucidate
some technical details. We adopt the metric $ds^2 >0$ for space-like
separations. 

\section{The action}\label{theactionsec}

Our starting point is the Henneaux-Teitelboim action \cite{HT,HTcile}  which
generalizes to curved backgrounds the Floreanini-Jackiw action \cite{FJ} 
for chiral bosons in two dimensions. On flat  background such  action gives
rise to the chirality condition
\begin{equation}
\partial_0\varphi + \partial_1 \varphi = 0
\end{equation}
provided some boundary conditions are satisfied \cite{FJ}. 
On curved background the
chirality condition becomes \cite{HT,BvNbook}
\begin{equation}
E_+^{\ \mu}\partial_\mu \varphi = 0
\end{equation}
where $E^\mu_+$ are inverse zweibeins.
In the following the key role will be played by the adimensional function
\begin{equation}
K= \frac{E_+^{\ 1}}{E_+^{\ 0}}=\frac{N}{\sqrt{h}}-N^1 = 
\frac{\sqrt{-g}-g_{01}}{g_{11}}.
\end{equation}
Right moving particles are described by $K>0$.
The action is provided by \cite{HT}
\begin{equation}
S= -\frac{1}{2}\int{d^2x ~\partial_1 \varphi (\partial_0 \varphi + 
K \partial_1 \varphi)}=\frac{1}{2}\int{d^2 x ~\varphi \partial_1(\partial_0+
K\partial_1)\varphi}.
\end{equation}
The variation of $S$ w.r.t $\varphi$ gives the equation of motion 
\begin{equation}
\partial_1(\partial_0+ K \partial_1)\varphi=0
\end{equation}
and the action vanishes on the equation of motion. 
Under a diffeomorphism the function $K$ undergoes the following passive
transformation
\begin{equation}
K(x)\rightarrow K'(x')=\frac{\frac{\partial {x'}^1}{\partial x^1}K(x) 
+\frac{\partial {x'}^1}{\partial x^0}}{\frac{\partial {x'}^0}{\partial
x^1}K(x) +\frac{\partial{x'}^0}{\partial x^0}}.
\end{equation}
More relevant in the following will be the active transformation
$K(x) \rightarrow K'(x)$ 
under which the action is invariant under diffeomorphisms \cite{BvNbook}, 
and that in the infinitesimal case takes the form 
\begin{equation}\label{deltakappa}
\delta_\xi \varphi = \Xi ~\partial_1\varphi~;~~~~~
\delta_\xi K = -\partial_0\Xi - \partial_1\Xi ~K + \Xi~\partial_1K
\end{equation}
where
\begin{equation}
\Xi = \xi^1-K\xi^0.
\end{equation}
On the equations of motion in computing the variation $\delta_\xi S$ 
we can ignore the variation  of the matter field $\varphi$ and we have 
\begin{eqnarray}
\delta_\xi S & = & -\frac{1}{2}\int{d^2x ~\partial_1 \varphi \delta_\xi K 
\partial_1 \varphi} \nonumber \\
               & = & -\frac{1}{2}\int{d^2x ~\delta_\xi g_{\mu \nu}
\frac{\partial K}{\partial g_{\mu \nu}}(\partial_1 \varphi)^2} \nonumber \\
               & = & -\int{d^2x ~\sqrt{-g}\nabla_\mu \xi_\nu 
\frac{1}{\sqrt{-g}}\frac{\partial K}{\partial g_{\mu \nu}}
(\partial_1 \varphi)^2}\nonumber \\
               & = & \int{d^2x ~\sqrt{-g} \xi_\nu \nabla_\mu \left(
\frac{1}{\sqrt{-g}}\frac{\partial K}{\partial g_{\mu \nu}}
(\partial_1 \varphi)^2\right)} 
\end{eqnarray}
from which we derive the classical energy momentum tensor
\begin{equation}
T^{\mu \nu}=-\frac{1}{\sqrt{-g}}\frac{\partial K}{\partial g_{\mu \nu}}
(\partial_1 \varphi)^2
\end{equation}
and we have
\begin{equation}
\nabla_\mu T^{\mu\nu}=0
\end{equation}
being $\nabla_\mu$ the usual covariant derivative.
An explicit calculation gives
\begin{equation}
\frac{\partial K}{\partial g_{\mu \nu}}=-\frac{1}{2}\sqrt{-g}k^\mu k^\nu
\end{equation}
with
\begin{equation}
k_1 =1;~~k_0= -K; \; \; k^\mu \equiv g^{\mu \nu}k_\nu.
\end{equation}
$T^{\mu\nu}$ is not a true tensor; on the other hand on the equations of motion
we can write
\begin{equation}
T^{\mu \nu}=\frac{1}{2}k^\mu k^\nu (\partial_1 \varphi)^2 = 
\frac{1}{2}\partial^\mu \varphi~ \partial^\nu \varphi
\end{equation}
and $T^{\mu \nu}$ becomes a true tensor.
\bigskip

\section{The gravitational anomaly via Pauli-Villars
regulators}\label{PVsection} 

To understand the meaning of the exact calculation it is
useful to make a few comments on the perturbative result. A perturbative
calculation was performed by use of the light-cone coordinates by Bastianelli
and van Nieuwenhuizen \cite{BvNpert,BvNbook}. Here we shortly describe the
equivalent 
calculation obtained by means of the PV regularization as it bears an
analogy to the exact calculation we shall perform in
Section \ref{exactanomaly}; moreover this technique can be extended also to
higher dimensions.

With $g_{\mu \nu}=\eta_{\mu \nu}+ h_{\mu \nu}$ we can write ${\cal L}= {\cal
L}_0 +{\cal L}_I $ where
\begin{equation}
{\cal L}_0 =-\frac{1}{2}\partial_1 \varphi (\partial_0 + \partial_1 )\varphi
\end{equation}
is the lagrangian for free chiral scalars and we have
\begin{equation}
{\cal L}_I = \frac{1}{2}\partial_1 \varphi \frac{h_{11}+h_{10}+h_{01}+
h_{00}}{2}\partial_1 \varphi =\frac{1}{2}\partial_1 \varphi h_{++} 
\partial_1 \varphi.
\end{equation}
The free propagator is given by
\begin{eqnarray}\label{freeprop} 
\langle T \varphi(x) \varphi(y) \rangle & = & i \left(\partial_1 
(\partial_0 + \partial_1 )\right)^{-1}\delta^2 (x-y)\nonumber \\
                                      & = & 
-i\int \frac{d^2 p}{(2\pi)^2}\sqrt{2}\frac{p_-}{p_1}
\frac{e^{ip\cdot(x-y)}}{(p^2-i\varepsilon)}
\end{eqnarray}
where $-i\varepsilon $ is the correct Feynman prescription.
\begin{eqnarray}
iW^{(2)}[h] & = & \frac{1}{2}\int{dx dy ~\langle 0 | T i{\cal L}_I (x)i
{\cal L}_I (y) |0\rangle} \nonumber \\
            & = & -\frac{1}{2}\int{d^2x d^2y ~\frac{1}{2}h_{++}(x)\langle
	    T\partial_1 \varphi(x)\partial_1 \varphi(x)\partial_1 
\varphi(y)\partial_1 \varphi(y)\rangle \frac{1}{2}h_{++}(y)} \nonumber \\
            & = & -\frac{1}{2}\int{d^2p ~h_{++}(p) U(p) h_{++}(-p)}
\end{eqnarray}
where we used the notation of \cite{AGW} 
\begin{equation}
h_{++}(x) = \frac{1}{2\pi}\int e^{ip\cdot x}h(p) d^2p
\end{equation}
and
\begin{equation}
x^\pm = \frac{x^1\pm x^0}{\sqrt{2}};~~~~p_\pm = \frac{p_1\pm p_0}{\sqrt{2}}
\end{equation}
\begin{equation}
U(p)=\frac{1}{4}\int{d^2 x ~e^{-ipx}\langle T\partial_1 
\varphi(x)\partial_1 \varphi(x)\partial_1 \varphi(0)\partial_1 \varphi(0)
\rangle}.
\end{equation}
Using the propagator (\ref{freeprop}) we obtain for $U(p)$ the divergent
expression 
\begin{equation}\label{U(p)}
U(p) = -\int{\frac{d^2k}{(2\pi)^2}(p+k)_1^2 \frac{(p+k)_-}{(p+k)_1
[(p+k)^2-i\varepsilon]}k_1^2 \frac{k_-}{k_1 [k^2-i\varepsilon]}}. 
\end{equation}
The pole $1/p_1$ in (\ref{freeprop}) is irrelevant in (\ref{U(p)}) due to the
vertex $p_1^2$ originating from $\partial_1\varphi \partial_1\varphi$ and 
as $ p^2 = 2 p_+ p_- $ the chiral propagator has a pole only 
for $p_+ =0$. We have  
\begin{equation}\label{propagator1}
\sqrt{2}\frac{p_-}{p_1 (p^2-i\varepsilon)}=
\sqrt{2}\frac{1}{2p_1 p_+ -i \frac{p_1}{p_-}\varepsilon}. 
\end{equation} 
On such pole we have $p_1 = -p_0 $ and as $\frac{p_1}{p_-}=\frac{1}
{\sqrt{2}}>0$, the previous propagator (\ref{propagator1}) is equivalent to 
\begin{equation}
\frac{1}{\sqrt{2}}\frac{1}{p_1 p_+ - i\varepsilon}. 
\end{equation} 
For completeness we introduce an IR regularization obtained by introducing a 
mass $m$ 
\begin{equation}\label{massiveprop}
\frac{1}{\sqrt{2}}\frac{1}{p_1 p_+ - i\varepsilon} \rightarrow
\frac{1}{\sqrt{2}}\frac{1}{p_1 p_+ + m^2 - i\varepsilon} \label{pcm}
\end{equation} 
while the PV regularization is obtained as usual by weighting
the one loop graphs with $m\rightarrow M_i$ with coefficients $c_i$ obeying
\cite{PV,anselmi}
\begin{eqnarray}\label{sumrules}
1+\sum_{i=1}^4 c_i=0~;~~m^2+\sum_{i=1}^4 c_i M^2_i=0;~~\nonumber \\ 
\log m^2+\sum_{i=1}^4 c_i \log M^2_i=0;
~~m^2 \log m^2+\sum_{i=1}^4 c_i M^2_i\log M^2_i=0
\end{eqnarray} 
where
$M_i \rightarrow \infty$ with $c_i$ not diverging in such a limit. This can be
achieved by setting $M_1^2=s~m^2$ , $M_2^2=s^2~ m^2$, $M_3^2=s^3 ~m^2$,
$M_4^2=s^4~ m^2$ with $s\rightarrow \infty$.  
At the end
we take the IR regulator $m$ to zero.  As usual \cite{BK} the IR regulated 
propagator (\ref{massiveprop}) 
with $m\neq 0$
does not describe any longer chiral bosons. $U$ becomes
\begin{equation}\label{Um} 
U(p,m^2)=
-\frac{1}{4}\int{\frac{d^2 k}{(2\pi)^2}\frac{(p+k)_1^2 k_1^2}{\left[(p+k)_1
(p+k)_+ + m^2 -i\varepsilon\right]\left[k_1 k_+ + m^2 -i\varepsilon \right]}}
\end{equation} 
and the regularized expression is
\begin{equation}\label{PVregularized} 
U_R(p)= U(p,m^2)+\sum_{i=1}^4 c_i U(p,M_i^2).
\end{equation}
Performing the following change of variables 
\begin{eqnarray}
l_1= k_1+ \frac{k_0}{2}; & {\cal P}_1 = p_1 + \frac{p_0}{2} \nonumber\\ 
l_0=\frac{k_0}{2}; & {\cal P}_0 = \frac{p_0}{2} 
\end{eqnarray} 
which is legal being (\ref{PVregularized}) convergent, we can rewrite the
amplitude (\ref{Um}) in the form 
\begin{eqnarray} 
U({\cal P}, \sqrt{2} ~m^2) & = & -4\int{\frac{d^2 l}{(2
\pi)^2}\frac{(l+{\cal P})_-^2 l_-^2}{\left[(l+{\cal P})^2 +
\sqrt{2}~m^2-i\varepsilon\right]\left[l^2 + \sqrt{2}~ m^2 - 
i\varepsilon\right]}}
\nonumber \\ & = & 2 T_{----} 
\end{eqnarray} 
where $ T_{----} $ is the not yet regulated amplitude
relative to non chiral  
scalar bosons. In Appendix A we compute the limit of the regularized amplitude
for $m^2\rightarrow 0$ and we have
\begin{eqnarray}\label{pertanom}
U_R(p) & = &
-\frac{i}{12\pi} \frac{{\cal P}_-^3}{{\cal P}_+} \nonumber \\ & = &
-\frac{i}{96\pi}\frac{(p_+ + p_-)^3}{p_+} =-\frac{i}{96\pi}(\frac{p_{-}^3}
{p_+}+3p_{-}^2+3p_{-}p_{+}+p_{+}^2). 
\end{eqnarray}
The last three terms in (\ref{pertanom}) can be eliminated by adding local
counter terms and (\ref{pertanom}) differs from the result of \cite{BvNbook} 
for chiral bosons and from the result of \cite{AGW} for chiral fermions by
similar local counter terms.
Thus we can write 
\begin{equation} 
W^{(2)}=\frac{1}{192\pi}\int{d^2 p
~h_{++}(p)\frac{p_-^3}{p_+}h_{++}(-p)}.
\end{equation} 
The variation under diffeomorfisms of $h_{++}$ is $\delta_\xi h_{++}(p) = 
2i p_+ \xi_+(p)$ and thus
\begin{equation}\label{pertanomaly} 
\delta_\xi W^{(2)} = -\frac{i}{48\pi}\int{d^2 p ~p_-^3
h_{++}(p)\xi_+(-p)}
\end{equation}
with 
\begin{equation}
\xi_\mu(p) = \frac{1}{2\pi}\int e^{-ip\cdot x}\xi_\mu(x) d^2x
\end{equation}
which is the same anomaly as the one found for chiral fermions.

We can also write using directly (\ref{pertanom}) 
\begin{equation}\label{pertanomaly2} 
\delta_\xi W^{(2)} = -\frac{i}{24\pi}\int{d^2 p ~p_1^3 h_{++}(p)
(\xi_1(-p)+\xi_0(-p)})
\end{equation}
which will bear a strong similarity with the exact result we shall obtain in
the following section.

\section{Exact calculation of the anomaly through the \break Schwinger- DeWitt
expansion}\label{exactanomaly} 
The generating functional is given by
\begin{eqnarray}\label{Z}
 Z[K] =e^{iW[K]}& = & 
\int{{\cal D}[\phi] \exp\left[-i\frac{1}{2}\int{d^2x\partial_1 
\phi (\partial_0 + K \partial_1) \phi }\right]}\nonumber\\
      & = & \int{{\cal D}[\phi] \exp\left[i\frac{1}{2}\int{d^2x \phi 
(\partial_1 \partial_0  + \partial_1 K \partial_1) \phi}\right]}\nonumber\\
      &\equiv & \left(\det -i(\partial_1 \partial_0  + 
\partial_1 K \partial_1  )\right)^{-\frac{1}{2}}.
\end{eqnarray}
As usual the
direct computation of (\ref{Z}) is difficult. 
However we shall be interested in
the variation of (\ref{Z}) under an infinitesimal diffeomorphisms 
which provides
us with the anomaly. 
We have
\begin{eqnarray}
i\delta_\xi W[K]= \int {\cal D}[\phi] e^{\frac{i}{2}\int
d^2x\phi(\partial_1(K\partial_1+\partial_0)\phi} \int
d^2x\frac{i}{2}\phi\partial_1(\delta_\xi K \partial_1\phi)/Z[K]\nonumber\\
=\frac{1}{2}\int d^2x ~\delta_\xi H ~G(x,x')|_{x'=x}
\end{eqnarray}
with
\begin{equation}
H= \partial_1(\partial_0 + K \partial_1)
\end{equation}
and
\begin{equation}
\delta_\xi H = \partial_1\delta K(x)\partial_1
\end{equation}
and $G(x,x')$ is the exact Green function in the external field $K$. We
regularize $G(x,x')$ \`a la Schwinger-DeWitt
\begin{equation}
G(x,x',\varepsilon)=i\langle x|\int_\varepsilon^\infty e^{iHt}dt|x'\rangle
\end{equation}
and thus
\begin{equation}\label{deltaW2}
i\delta_\xi W[K]=\frac{i}{2}\int_\varepsilon^\infty dt\int d^2x 
\delta_\xi H \langle x|e^{itH}
|x'\rangle|_{x'=x}.
\end{equation}
We exploit now the fact that
\begin{equation}
\delta_\xi H = \partial_1 (\Xi~ H) - H~ \Xi \partial_1
\end{equation}
with $\Xi = \xi^1-K\xi^0$
and take advantage of the equation satisfied by  
$\langle x|e^{iH}|x'\rangle$ due to the evolution equation
\begin{equation}
\frac{de^{iHt}}{dt}=iH e^{iHt}=ie^{iHt} H~,
\end{equation}
to rewrite eq.(\ref{deltaW2}) as
\begin{eqnarray}
\delta_\xi W  =  \frac{i}{2}\int_\varepsilon^\infty dt 
\frac{d}{dt}\int d^2x d^2x' ~\delta(x-x') \Xi(x') \left(\partial'_1 + 
\partial_1\right) \langle x|e^{iHt}|x'\rangle=\nonumber \\
-\frac{i}{2}\int d^2x d^2x' ~\delta(x-x') \Xi (x') \left(\partial'_1 + 
\partial_1\right)\langle x|e^{iH\varepsilon}|x'\rangle.
\end{eqnarray}
We compute the short time behavior of $\langle x|e^{iHt}|x'\rangle$ 
by using the Schwinger-DeWitt technique. The operator 
\begin{equation}
H = \partial_1 K\partial_1 + \partial_1\partial_0
\end{equation}
is the Laplace-Beltrami operator in the metric 
\begin{equation}\label{metric}
g_{11}= 0 ;~~~~g_{10}=g_{10}=2;~~~~g_{00}= -4 K
\end{equation}
for which $\sqrt{-g}=2$. We apply  the well known expansion
\cite{dewitt,seeley} 
\begin{equation}
\left\langle x\right|e^{itH}\left|x'\right\rangle=
\frac{(-g(x))^{\frac{1}{4}}\Delta^{\frac{1}{2}}(x,x')
(-g(x'))^{\frac{1}{4}}}{4\pi t}
e^{\frac{i\sigma(x,x')}{2t}}\sum_{i=1}^\infty (it)^n a_n(x,x')
\end{equation}
where $2\sigma(x,x')$ is the square of the geodesic distance between $x$ and
$x'$ 
and $D(x,x') = (-g(x))^{\frac{1}{2}}\Delta(x,x')
(-g(x'))^{\frac{1}{2}}$ is the Van Vleck-Morette determinant i.e.
\begin{equation}
\det\left[\frac{\partial^2\sigma(x,x')}{\partial x^\mu\partial x'^\nu}\right].
\end{equation}
Using \cite{BV}
\begin{equation}
a_0(x,x')=1;~~~\partial_1\Delta(x,x')|_{x'=x}=0;~~~~
\partial_1 a_1(x,x')|_{x'=x} =\frac{1}{12}\partial_1 R(x)
\end{equation}
being $R(x)$ the scalar curvature of the metric (\ref{metric}), we obtain
\begin{equation}
\partial_1\langle x|e^{iHt}|x'\rangle = \frac{i}{24\pi}\partial_1R.
\end{equation}
A simple calculation of the curvature of the metric (\ref{metric}) gives
\begin{equation}
R= -\partial_1^2K
\end{equation}
and thus
\begin{equation}\label{nonpert}
\delta_\xi W = -\frac{1}{24\pi}\int d^2x~\Xi(x)~\partial_1^3K(x)=
\frac{1}{24\pi}\int d^2x~\partial_1^3\Xi(x) ~K(x)\equiv G^E[K,\Xi] 
\end{equation}
which is the Einstein anomaly.
This is the non perturbative result and it agrees with  the general  form of
the anomaly for conformally coupled scalar matter with gravity \cite{BTvP2}. 
As for weak
gravitational fields we have 
\begin{equation}
K = 1 - h_{++}
\end{equation}
eq.(\ref{nonpert}) agrees with the weak external field result
(\ref{pertanomaly2}) found through PV regularization.

\section{\bf Wess-Zumino consistency condition and non triviality of the 
anomaly}
It is useful to introduce the anticommuting diffeomorphism ghosts $v^1, v^2$.
Using eq.(\ref{deltakappa}) we can write the BRST variation
\begin{equation}\label{deltaK}
\delta K= -\partial_0 V-\partial_1 V K + V\partial_1 K
\end{equation}
where $V= v^1-Kv^0$. As $K$ is the only function appearing in the theory we
see that all diffeomorphisms are described by the single ghost $V$. Using
\begin{equation}\label{deltav}
\delta v^\mu =v^\lambda\partial_\lambda v^\mu
\end{equation}
and eq.(\ref{deltaK})
it is easily proved that
\begin{equation}\label{deltaV}
\delta V=V\partial_1V.
\end{equation}
The algebra of diffeomorphisms requires
\begin{equation}
\delta^2 K =0
\end{equation}
a relation which can be explicitely verified using
eqs.(\ref{deltaK},\ref{deltaV}).
We can now verify the Wess-Zumino consistency condition for the found anomaly
(\ref{nonpert}). In fact we have
\begin{equation}
\delta(\partial_1^3 V K)= \partial_1^3 (V\partial_1 V)K-\partial_1^3V\delta K=
\partial_1(V\partial_1^3VK)+\partial_1^3V\partial_0V.
\end{equation}
The first term on the r.h.s. is a divergence while for the second, integrating
by parts and 
using the anticommutativity of $V$, we have
\begin{eqnarray}
\delta G^E[K, V] & = & {\rm const.} ~\int{d^2x ~\partial_1^3V\partial_0V}
\nonumber \\
          & =       & {\rm const.} ~\int{d^2x ~\partial_0 V \partial_1^3 V}
\nonumber \\
          & =       & -{\rm const.} ~\int{d^2x ~\partial_1^3V\partial_0V}=0.
\end{eqnarray}
The Wess-Zumino consistency relation can also be written as
\begin{equation}\label{WZ}
\delta Q^1_2 = - dQ^2_1
\end{equation}
where $Q^1_2= V\partial_1^3K$ and $Q^2_1$ is a $1$-form of degree $2$ in the
ghost $V$.

Our aim now will be the following: construct the descending cohomology chain
and show algebraically by examining the last term of the chain, 
that the found anomaly 
(\ref{nonpert}) is
non trivial. Then prove the uniqueness of the last term and also prove that
the ascent of the cohomology chain is unique, thus proving on purely algebraic
ground that the found non perturbative anomaly (\ref{nonpert}) is unique up to
a multiplicative factor.
We write
\begin{equation}
\delta(V\partial_1^3 K dx^1 \wedge dx^0 ) = -dQ^2_1.
\end{equation}
The $1$-form $Q^2_1$ can be explicitely found
\begin{eqnarray}
Q^2_1 & = & -\frac{1}{2}(\partial_1^3 V V)dx^1
-\left[\frac{1}{2}\partial_1^2V\partial_0V-\frac{1}{2}\partial_1V
\partial_1\partial_0V+\frac{1}{2}V
\partial_1^2\partial_0V \right. \nonumber \\
      &   & \mbox{} -\partial_1V\partial^2_1VK-V\partial_1V\partial_1^2K+ 
V\partial_1(\partial_1^2V K)\biggr]dx^0.
\end{eqnarray}
The fact that such a form $Q^2_1$ satisfying (\ref{WZ}) can be constructed is 
the content of  
the Wess- Zumino consistency condition.
Using the algebraic Poincar\'e lemma \cite{stora1,stora2,cotta,wald,BDK1} 
a $0$- form $Q^3_0$
must exist such that 
\begin{equation}\label{Q21chain}
\delta Q^2_1 = -dQ^3_0.
\end{equation}
Such form is easily found
\begin{equation}
Q^3_0 = \frac{1}{2}V\partial_1V\partial_1^2V. 
\end{equation}
It is easily proved that if $Q^1_2$ is a trivial anomaly i.e.
\begin{equation}
Q^1_2 = \delta X^0_2+d X^1_1
\end{equation}
it follows
\begin{equation}\label{deltaQ30}
Q^3_0 = \delta X^2_0.
\end{equation}
Thus if we show that no $X^2_0$ exists satisfying (\ref{deltaQ30}) we prove
algebraically the non triviality of $Q^1_2$. One notices that $X^2_0$ must
contain two derivatives and thus the most general $X^2_0$ is given by
\begin{eqnarray}\label{X20}
X^2_0 & = & V\partial_1^2 V g_1+V\partial_1\partial_0V
g_2+V\partial_0^2 V g_3+ \partial_1V\partial_0V g_4+
 \nonumber \\
      &   & 
+V\partial_1 V \partial_1Kg_5+V\partial_1 V \partial_0K g_6
+ V\partial_0 V \partial_1K g_7+V\partial_0V \partial_0K g_8.
\end{eqnarray}
One notices also that $V\partial_1V\partial_1^2V$ can originate only from 
the $g_1$ and $g_5$ terms. Thus we must have
\begin{equation}
Kg_1'(K)-Kg_5(K)=1.
\end{equation}
Furthermore as $V\partial_0V\partial_0^2V$ can originate only from the terms
$g_3$ and $g_8$ we have the condition
\begin{equation}
g_3'(K)=g_8(K)
\end{equation}
which allows us to write the variation $\delta X^2_0$ 
in terms of the parameter $s=2g_3+Kg_3'$. 
One
also notices that the term proportional to 
$V\partial_1V\partial_0V \partial_0K$ obtained from the variation of the above
two terms can be eliminated only by the variation of the $g_6$ term giving rise
to the relations
\begin{equation}
s'(K)=g_6'(K)~~~~{\rm i.e.} ~~~~g_6(K)=s(K)+c
\end{equation}
where $c$ is a constant. Putting now all the variations together we obtain the
following relations given by the vanishing of the coefficients of
$V\partial_1V\partial_0V\partial_1K$, 
$V\partial_1V\partial_1\partial_0V$,
$V\partial_0V\partial_1\partial_0 V$,  
$V\partial_0V\partial_1^2 V$,
$V\partial_1V\partial_0^2V$,   
\begin{eqnarray}
0 & = & 2g_7(K)+s(K)+c+Kg_7'(K)+g_4'(K)-1/K^2-g_1''(K)\nonumber\\
0 & = & 1/K+g_4(K)-Ks(K)-Kc+g_2(K)-g_1'(K)+Kg_2'(K)\nonumber\\
0 & = & -g_7(K)-s(K)+g_2'(K)\nonumber\\
0 & = & -Kg_7(K)-g_4(K)-g_2(K)+g_1'(K) \label{fprimo}\nonumber\\
0 & = & c 
\end{eqnarray}
Using simple algebra we arrive to the obstruction $0=1/K$ 
which proves algebraically the non triviality of the found anomaly.

\section{\bf Cohomological derivation of the anomaly}\label{cohosection}

In the following discussion we shall denote by $S^n(m)$ the space of terms
containing $n$ ghosts and $m$ derivatives, e.g. $V\partial_1^2V \partial_0 K
f(K)\in S^2(3)$.
 
In the present section we shall give a very simple cohomological treatment of
the non perturbative anomaly based on the fact that a single ghost describes
all the diffeomorfisms in each chirality sector \cite{becchi}. 
First we prove that the last
term $Q^3_0$ in the cohomological chain 
is unique apart for the addition of a trivial term i.e. a $\delta$ variation,
and then, starting from such $Q^3_0$ we prove that the cohomology chain can be
climbed up in a unique way leading, in a pure algebraic way to the result
(\ref{nonpert}) apart a multiplicative constant.

The first question is equivalent to the
cohomological problem of proving that the sequence
\begin{equation}\label{S00sequence}
S^0(0) \stackrel{\delta}{\rightarrow} S^1(1) 
\stackrel{\delta}{\rightarrow} S^2(2) \stackrel{\delta}{\rightarrow} 
S^3(3) \stackrel{\delta}{\rightarrow} S^4(4) 
\stackrel{\delta}{\rightarrow} 0
\end{equation}
differs from an exact sequence only in the penultimate junction due to the
presence of the non trivial term $N^3_0\equiv {\rm const.}~
V\partial_1V\partial^2_1V$. 

It will be useful in the present section to perform a change of basis by
replacing the basis element $\partial_0V$ by $W\equiv \delta K$, which is
equivalent to it due to the relation (\ref{deltaK}). Then the algebra we
shall use is simply
\begin{equation}\label{simplealgebra}
\delta V=V\partial_1V;~~~~\delta K=W;~~~~\delta W=0
\end{equation}
and that will be sufficient to perform all calculations.

$S^0(0)$ is the space of the functions
$f(K)$ and $\delta S^0(0)\subset S^1(1)$ has also dimension $1$, $\delta
f(K)=Wf'(K)$ . The space $S^1(1)$ has dimension $4$ and its elements can be
written as 
\begin{equation}\label{S11space}
\partial_1 V h_1+W h_2+V\partial_1 K h_3+V\partial_0 K h_4.
\end{equation}

In examining the kernel of $\delta$ from $S^1(1)$ into $S^2(2)$ we can gauge
fix to zero $h_2$ to zero by adding the variation of $f(K)$ with
$f'(K)=-h_2(K)$. It 
is then easily shown that the kernel on the remaining space is
$h_1=h_3=h_4=0$ as
\begin{eqnarray}\label{deltaS11}
\delta(\partial_1V h_1)=V\partial_1^2V h_1+\dots \nonumber\\
\delta(V\partial_1K h_3)=V\partial_1 V \partial_1 Kh_3+\dots\nonumber\\
\delta(V\partial_0K h_4)=V\partial_1 V \partial_0 Kh_4+\dots
\end{eqnarray}
and the terms written explicitely in each equation have no counterpart
in the remaining two 
equations.
This proves the exactness of the first short sequence. The
space $S^2(2)$ has dimension $8$ and using (\ref{deltaS11}) we can gauge fix
to zero the $3$ terms
$V\partial_1^2Vg_1,~V\partial_1V\partial_1Kg_2,\break 
V\partial_1V\partial_0Kg_3$
leaving the $5$ terms $\partial_1V W g_4,~VW\partial_iKg^{(i)}_5,~
V\partial_iWg^{(i)}_6$, $(i=1,0)$ .
 
For the variations we have
\begin{eqnarray}\label{deltaS22}
\delta(\partial_1V Wg_4)=V\partial_1^2V Wg_4 \nonumber\\
\delta(VW\partial_iK g_5^{(i)})=V\partial_1 V W\partial_i Kg_5^{(i)}+
\dots\nonumber\\
\delta(V\partial_iW  g_6^{(i)})=V\partial_1 V \partial_iW g_6^{(i)}+\dots
\end{eqnarray}
 and again the written terms are unique in the variations so that we obtain for
 the kernel $g_4=g_5^{(i)}=g_6^{(i)}=0$.
Finally $S^3(3)$ has dimension $8$ and gauge fixing to zero
$V\partial_1^2V Wf_4,~V\partial_1 V W\partial_i Kf_5^{(i)} ,V\partial_1 V
\partial_iW f_6^{(i)}$ we are left with the terms $V\partial_1 V\partial_1^2V
f_1$, \break $VW\partial_1W f_2$, $ VW \partial_0 W f_3$ 
from which, as $\delta(V\partial_1V\partial_1^2V)=0$, we obtain for the kernel
$f'_1(K)=0,~ f_2(K)=f_3(K)=0$. 

Thus we found that the most general solution of $\delta Q^3_0=0$ can be
written in the form
\begin{equation}
Q^3_0 = N^3_0 + \delta X^2_0
\end{equation}
where
\begin{equation}
N^3_0 = {\rm const.}V\partial_1V \partial_1^2 V~,
\end{equation}
\bigskip
\begin{figure}
\begin{center}
\epsffile{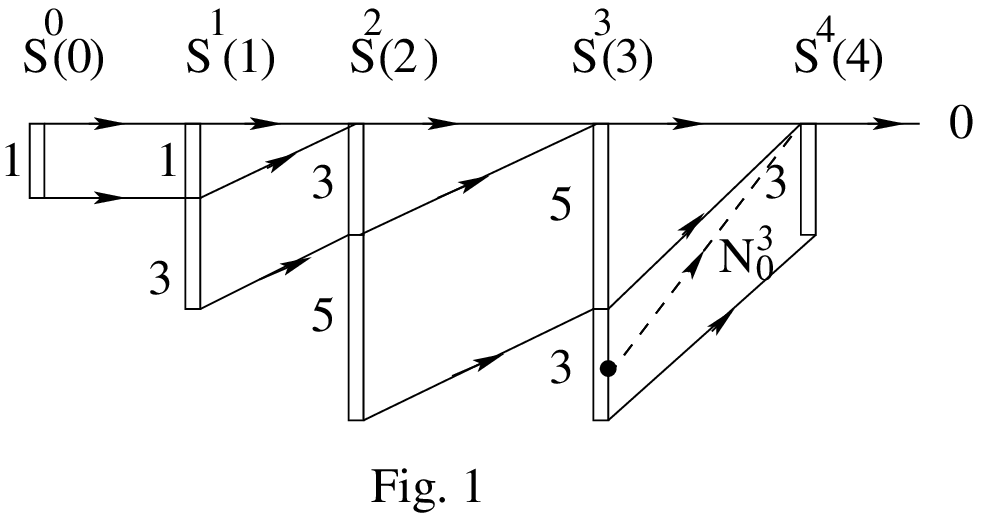}
\end{center}
\end{figure} 

i.e the short sequence around $S^3(3)$ in (\ref{S00sequence}) 
is exact except for the element $N^3_0$. 
Thus the last term $Q^3_0$ of the cohomological
sequence is unique up to a  multiplicative constant and trivial additions. 
The sequence (\ref{S00sequence}) is shown in Fig.1 where the numbers
on the vertical bars denote the dimension of the space.

The next step is to show the uniqueness of the term $Q^2_1$ of
eq.(\ref{Q21chain}) up to trivial additions.
To this end we have to prove that the kernel of $\delta$ from
$S^2(3)$ into $S^3(4)$ is zero, modulo the trivial terms
$\delta S^1(2)$. 
To this end we shall show that the sequence
\begin{equation}\label{S01sequence}
0 \stackrel{\delta}{\rightarrow} S^0(1) \stackrel{\delta}{\rightarrow} 
S^1(2) \stackrel{\delta}{\rightarrow} S^2(3) 
\stackrel{\delta}{\rightarrow} S^3(4)
\end{equation}
is exact.  $S^0(1)$ has only two elements, $\partial_1K ~h_1(K)$ and
$\partial_0 K ~ h_2(K)$ and it is immediate that the kernel of $\delta$ from
$S^0(1)$ in $S^1(2)$ is the zero. There are $13$ elements in $S^1(2)$ two of
which e.g. $\partial_iWf^{(i)}$ can be gauge fixed to zero. An elementary
calculation reported in Appendix B shows that the kernel of such gauge fixed
$11$ dimensional space is trivial, that we can operate $11$ gauge fixing in
$S^2(3)$ and that the kernel of $\delta$ acting on such gauge fixed space is
trivial. 

The sequence (\ref{S01sequence}) is depicted in Fig.2. 

We are left now with climbing the last step of the cohomology chain i.e. after
proving the uniqueness, up to trivial terms, of $Q^2_1$ we want to 
prove the uniqueness, up to trivial terms, of the solution of
\begin{equation}
\delta Q^1_2= -d Q^2_1
\end{equation}
of which we know already a solution i.e. $Q^1_2 = {\rm const.}~
V\partial_1^3 K dx^1\wedge dx^0$. Thus we
want to show that if 
\begin{equation}
\delta Q^1_2=0
\end{equation}
we have $Q^1_2=\delta X^0_2$. It corresponds to proving the exactness of 
sequence 
\begin{equation}\label{S02sequence}
0 \stackrel{\delta}{\rightarrow} S^0(2) \stackrel{\delta}{\rightarrow} 
S^1(3) \stackrel{\delta}{\rightarrow} S^2(4). 
\end{equation}
\begin{figure}
\begin{center}
\epsffile{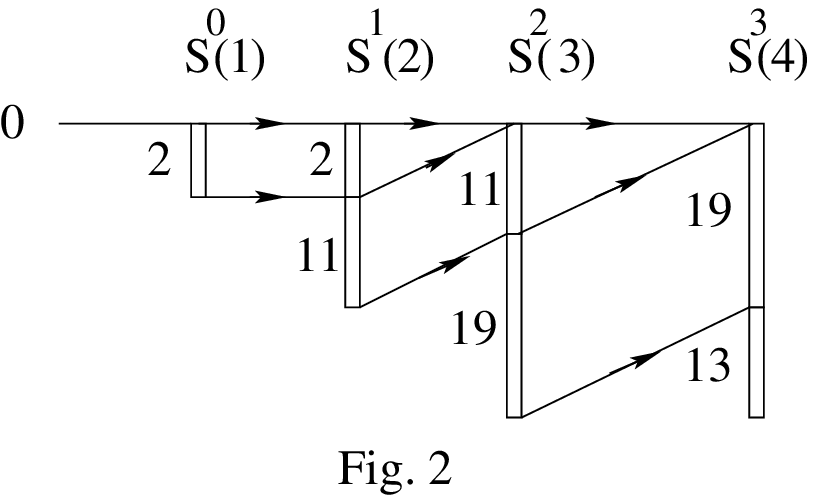}
\end{center}
\end{figure}
The space $S^0(2)$ has dimension $6$, $S^1(3)$ has dimension $36$
and $S^2(4)$ dimension $88$. The dimension of $\delta S^0(2)$ is
$6$ which means that we can operate $6$ gauge fixings in $S^1(3)$
corresponding to fixing to zero the coefficients of the six terms
$\partial_1W \partial_1K f$, $\partial_0W \partial_0K f$,
$\partial_0W \partial_1 K f$, $\partial_1^2W f$, $\partial_0^2W f$,
$\partial_1\partial_0W f$, where $f$ denote arbitrary functions of $K$.
\begin{figure}
\begin{center}
\epsffile{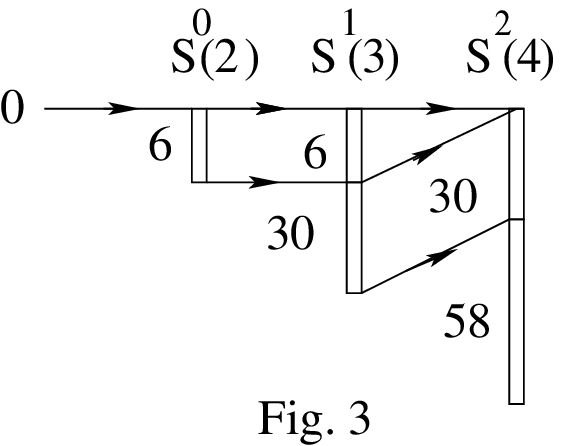}
\end{center}
\end{figure}
An elementary calculation reported in Appendix B proves that on such
gauge fixed space the only solution of $\delta X=0$, for $X\in S^1(3)$, is
zero. Sequence (\ref{S02sequence}) is shown in Fig.3.

\section{Conclusions}
In the present paper we developed two direct derivations of the consistent
gravitational anomaly generated by a chiral boson in two dimensions. After
revisiting by means of the PV regularization the perturbative result, we give
two 
independent derivations of the exact consistent gravitational anomaly. The
first is based on the Schwinger-DeWitt technique and the other on
cohomological methods. The simplicity of the cohomological method is due to
the fact that only one function of the metric appears in the
Henneaux-Teitelboim action and that a single ghost generates all the
diffeomorphisms. In such an approach we show the uniqueness of the final term
in the cohomological chain and that the cohomological chain can be climbed up
in a unique way. This is due to the exactness of three cohomological sequences
but for a junction which gives rise to the anomaly.

\bigskip\bigskip

\section*{Acknowledgments}
We are grateful to Damiano Anselmi for useful discussions.

\section*{Appendix A}

In this Appendix we report the explicit formulae of the PV regularized
calculation which have been used in Section \ref{PVsection} 
to find the perturbative anomaly.
The non chiral amplitude is usually \cite{BK} decomposed as

\begin{eqnarray}\label{BKamplitude}
T_{\mu \nu \rho \sigma}(p) & = & p_\mu p_\nu p_\rho p_\sigma T^v_1(p^2) + 
\left(p_\mu p_\nu g_{\rho \sigma} + p_\rho p_\sigma g_{\mu \nu} \right) 
T^v_2(p^2)\nonumber \\
                           &   & \mbox{}+ 
\left(p_\mu p_\rho g_{\nu \sigma} + 
p_\mu p_\sigma g_{\nu \rho} + p_\nu p_\rho g_{\mu \sigma} + 
p_\nu p_\sigma g_{\mu \rho} \right) T^v_3 (p^2)\nonumber \\
                           &   & \mbox{}+ g_{\mu \nu}g_{\rho \sigma}
T^v_4(p^2) + \left(g_{\mu \rho}g_{\nu \sigma} + 
g_{\mu \sigma}g_{\nu \rho}\right) T^v_5(p^2).
\end{eqnarray}
but due to the existence in dimension $2$ of the identically zero tensor
\cite{BK,anselmipc}
\begin{eqnarray}
& \left(p_\mu p_\nu g_{\rho \sigma} + p_\rho p_\sigma g_{\mu
\nu}\right)-\frac{1}{2}\left(p_\mu p_\rho g_{\nu \sigma} + p_\mu p_\sigma
g_{\nu \rho} 
+ p_\nu p_\rho g_{\mu \sigma} + p_\nu p_\sigma g_{\mu \rho}\right)\nonumber \\
& \mbox{}
- p^2 g_{\mu \nu}g_{\rho \sigma} + \frac{p^2}{2}
\left(g_{\mu \rho}g_{\nu \sigma} + g_{\mu \sigma}g_{\nu \rho}\right)
\end{eqnarray}
it is possible to reduce the invariants from $5$ to $4$ 
\begin{eqnarray}\label{ouramplitude}
T_{\mu \nu \rho \sigma}(p) & = & p_\mu p_\nu p_\rho p_\sigma T_1(p^2) + 
\left(p_\mu p_\nu g_{\rho \sigma} + p_\rho p_\sigma g_{\mu \nu} \right) 
T_2(p^2)\nonumber \\
                           &   & \mbox{}+ g_{\mu \nu}g_{\rho \sigma}
T_4(p^2) + \left(g_{\mu \rho}g_{\nu \sigma} + 
g_{\mu \sigma}g_{\nu \rho}\right) T_5(p^2).
\end{eqnarray}
As the PV regularization does not need any
continuation in dimensions we can work with the form (\ref{ouramplitude}).

The relation between our invariants and those of eq.(\ref{BKamplitude}) is
\begin{eqnarray}
T_1 = T^v_1;~~~~T_2 = T^v_2+2 T^v_3 ;~~~~T_3 =
T^v_3- T^v_3=0\nonumber\\
T_4 = T^v_4-2p^2 T^v_3;~~~~ T_5 = T^v_5+p^2 T^v_3.
\end{eqnarray}
We obtain regulating as in eqs.(\ref{sumrules},\ref{PVregularized}), in the
limit $s\rightarrow \infty$
\begin{eqnarray}\label{ourPV}
T_1 (p^2) & = & -\frac{i}{2\pi}\int_{-\frac{1}{2}}^\frac{1}{2}
{d\beta \frac{\left(\frac{1}{4}-\beta^2\right)^2}
{p^2 \left(\frac{1}{4}-\beta^2\right) + m^2}} \nonumber\\
T_2 (p^2) & = & \frac{i}{2\pi}\int_{-\frac{1}{2}}^{\frac{1}{2}}d\beta \left
\{
\frac{p^2\left(\frac{1}{4}-\beta^2\right)^2}
{p^2\left(\frac{1}{4}-\beta^2\right) + m^2}\right\} \label{T2b}= -p^2
T_1\nonumber\\ 
T_4 (p^2) & = & \frac{i}{8\pi}\int_{-\frac{1}{2}}^{\frac{1}{2}} 
d\beta \left\{ 
\left[p^2\left(\frac{1}{4} - \beta^2\right) + 
m^2\right] \ln\left[1 + \frac{p^2}{m^2}\left(\frac{1}{4}
-\beta^2\right)\right] \right. \nonumber \nonumber\\
& & \left. 
- 4p^2\left(\frac{1}{4}-\beta^2\right) - 
\frac{\left[p^2\left(\frac{1}{4}-\beta^2\right) - 
m^2 \right]^2}{p^2\left(\frac{1}{4}-\beta^2\right)+ m^2} + 
m^2 \right\}= (p^2)^2 T_1 \nonumber\\
T_5 (p^2) & = &
- \frac{i}{8\pi}\int_{-\frac{1}{2}}^\frac{1}{2}d\beta 
\left[\left(p^2\left(\frac{1}{4}-3\beta^2\right)+ m^2\right) 
\ln \left(1 + \frac{p^2}{m^2}\left(\frac{1}{4}-
\beta^2\right)\right) \right.\nonumber \\
                &   & \left. \mbox{} -p^2 
\left(\frac{1}{4}-\beta^2\right)\right]=0 \label{T5b}.
\end{eqnarray}
where we used
\begin{equation}
\int^{\frac{1}{2}}_{-\frac{1}{2}}d\beta \left(x(\frac{1}{4}-3
\beta^2)+1\right)\ln\left(1+x(\frac{1}{4}-\beta^2)\right)=\frac{x}{6}. 
\end{equation}
The amplitudes $T_i$ satisfy the diffeomorphism Ward identities
\begin{equation}
p^2 T_1+ T_2+ 2T_3=0;~~~~p^2 T_2+T_4=0;~~~~ p^2 T_3+T_5=0.	
\end{equation}
The integral appearing in eq.(\ref{ourPV}) can be explicitely computed, 
but in order to take the 
limit $m^2\rightarrow 0$ is better to keep the form (\ref{ourPV}). All these 
amplitudes are finite as the infrared regulator $m^2$ goes to zero
with the results
\begin{equation}
T_1 = -\frac{i}{12 \pi p^2};~~~~T_2=\frac{i}{12 \pi};~~~~
T_4=-\frac{i p^2}{12 \pi} ;~~~~T_5=0. 
\end{equation}
\bigskip

\section*{Appendix B}
In the present appendix we give the details in the analysis of the
exactness of the sequences of Fig.2,3. The calculations are trivial but for
clearness we report the details.

With regard to sequence (\ref{S01sequence}) the variation of the $11$ terms
remaining after the gauge fixing are, using (\ref{simplealgebra})
\begin{eqnarray}
\delta(V\partial_i\partial_jK f) &=&V\partial_1V \partial_i\partial_j K f+\dots
\nonumber\\
\delta(V\partial_i K\partial K_j f) &=&V\partial_1V \partial_i K\partial_j K
f+\dots \nonumber\\
\delta(\partial_1 V\partial_i K f) &=&V\partial_1^2V \partial_iK
f+\dots\nonumber\\ 
\delta(\partial_1^2 V f) &=&V\partial_1^3 V f+\dots\nonumber\\
\delta(W\partial_i K f) &=&-W\partial_iW f
\end{eqnarray}
where $f$ denotes functions of $K$. 
All the explicitely written terms on the r.h.s. appear only once in the
variation and this proves that the kernel from $S^1(2)$ to $S^2(3)$ is trivial
and at the same time that we can perform $11$ gauge fixings corresponding to
the explicitely written terms. The variation of the remaining $19$ terms are,
using (\ref{simplealgebra})
\begin{eqnarray}
\delta(V W \partial_i\partial_jK f) &=& V\partial_1V  W \partial_i\partial_jK
f+\dots \nonumber\\
\delta(V W \partial_iK\partial_jK f) &=& V\partial_1V  W \partial_iK\partial_jK
f+\dots \nonumber\\
\delta(V \partial_iW\partial_jK f) &=& V\partial_1V \partial_i W\partial_jK
f+\dots \nonumber\\
\delta(V \partial_i\partial_jW  f) &=&V\partial_1 V \partial_i\partial_jW f
+\dots \nonumber\\
\delta(\partial_1 V W\partial_jK f) &=&V\partial_1^2 V W\partial_jK f
+\dots \nonumber\\
\delta(\partial_1 V \partial_iW  f) &=&V\partial_1^2 V
\partial_iW f +\dots \nonumber\\
\delta(\partial^2_1 V W  f) &=&V\partial_1^3 V W f
+\dots \nonumber\\
\delta(\partial_1V \partial_1^2 V  f) &=&V \partial_1 V\partial_1^3 V f+
\dots \nonumber
\end{eqnarray}
As all the reported terms on the r.h.s. appear only once in the variations,
the kernel of $\delta$ from $S^2(3)$ into $S^3(4)$ is trivial.

\bigskip

With regard to the sequence (\ref{S02sequence}), in $S^1(3)$ we gauge fix to
zero the six terms  $\partial_i\partial_j W f$, $\partial_iW\partial_i K f$,
$\partial_0W \partial_1Kf$ $(i,j=1,0)$. To prove that the kernel of $\delta$
on the surviving $30$ dimensional space is zero one notices as above, that for
the $23$ terms containing $V$ the results obtained by varying simply $V$ are
all independent. For the remaining $7$ terms we have
\begin{eqnarray}
\delta(W\partial_i\partial_jK f) &=& -W \partial_i\partial_jW f \nonumber\\
\delta(W\partial_iK\partial_jK f) &=& -W \partial_iW\partial_jK f-W
\partial_iK\partial_jW f 
\nonumber\\
\delta(\partial_1W\partial_0Kf) &=&-\partial_1W \partial_0Wf-\partial_1W
\partial_0 K W f' 
\end{eqnarray}
and due to the independence of the terms on the r.h.s. the kernel from
$S^1(3)$ to $S^2(4)$ is trivial.  

\eject

\end{document}